\documentclass[twocolumn,showpacs,preprintnumbers,amsmath,amssymb]{revtex4}

\usepackage{graphicx}
\usepackage{dcolumn}
\usepackage{bm}
\usepackage{ulem}
\begin{document}

\preprint{APS/123-QED}

\title{Fluid-like dissipation of magnetic turbulence at electron scales in the solar wind}

\author{O. Alexandrova}
\email{olga.alexandrova@obspm.fr}
\altaffiliation{LESIA-Observatoire de Paris, CNRS, UPMC Universit\'e Paris 06, Universit\'e Paris-Diderot, 5 place J.~Janssen, 92190 Meudon, France.}
\author{C. Lacombe}
\affiliation{LESIA-Observatoire de Paris, CNRS, UPMC Universit\'e Paris 06, Universit\'e Paris-Diderot, 5 place J.~Janssen, 92190 Meudon, France.}
\author{A. Mangeney}
\affiliation{LESIA-Observatoire de Paris, CNRS, UPMC Universit\'e Paris 06, Universit\'e Paris-Diderot, 5 place J.~Janssen, 92190 Meudon, France.}
\author{R. Grappin}
\affiliation{LUTH-Observatoire de Paris, CNRS, Universit\'e Paris-Diderot, 5 place J.~Janssen, 92190 Meudon, France \&  LPP, Ecole Polytechnique 91128 Palaiseau, France.}

\date{October 28, 2011}

\begin{abstract}

The turbulent spectrum of magnetic fluctuations in the solar wind displays a spectral break at ion characteristic scales.  At electron scales the spectral shape is not yet completely established. 
 Here, we perform a statistical study of 102 spectra at plasma kinetic scales, measured by the Cluster/STAFF instrument  in the free solar wind. 
We show that the magnetic spectrum in the high frequency range, $[1,400]$~Hz, has a form  similar to what is found in hydrodynamics in the dissipation range $\sim A k^{-\alpha}\exp{(-k\ell_d)}$.  The dissipation scale  $\ell_d$ is found to be correlated with the electron Larmor radius  $\rho_e$.  The spectral index $\alpha$ varies in the range $[2.2,2.9]$ and is anti-correlated with $\ell_d$, as expected in the case of the balance between the energy injection and the energy dissipation. The coefficient $A$ is found to be proportional to the ion temperature anisotropy, suggesting that local ion instabilities may play some r{\^o}le  for the solar wind turbulence at plasma kinetic scales. The exponential spectral shape  found here indicates that the effective dissipation of magnetic fluctuations in the solar wind has a wave number dependence similar to that of the resistive term in collisional fluids   $\sim  \triangle \delta B \sim k^2 \delta B$.

\end{abstract}

\maketitle

In ordinary fluids, turbulent fluctuations are unpredictable, but their statistics are predictable and universal \citep{Frisch95}; turbulent spectra follow the power-law $\sim k^{-5/3}$ for any local conditions ($k$ being the wave number). This empirical result was explained by Kolmogorov \citep{Kolmogorov1941} assuming self similarity of turbulent fluctuations between the energy injection scale (the largest scale of the system) and the dissipation one $\ell_d$ (the smallest scale).

In the magnetized solar wind, collisions are very rare (the mean free path is of the order of 1~AU), the dissipation process at work and  the characteristic dissipation length are not known  precisely. Moreover, in a magnetized plasma, it is difficult to imagine self-similarity at all scales where turbulent fluctuations are observed, since there exist several spatial and temporal characteristic scales, such as the ion Larmor radius $\rho_{i}=\sqrt{2kT_{i\perp}/m_i}/(2\pi f_{ci})$,  the ion inertia length $\lambda_{i}=c/\omega_{pi}$,  the corresponding electron scales $\rho_{e}, \lambda_{e}$, and the ion and electron cyclotron frequencies $f_{ci}$, $f_{ce}$. At these scales,  the dominant physical processes change, which affects the scaling of the energy transfer time and furthermore the energy transfer rate,  leading to spectral shape changes.

In this Letter we study magnetic field turbulent fluctuations at plasma kinetic scales, starting at ion scales and going beyond electron spatial scales.

The broad solar wind spectra ranging from magnetohydrodynamic scales (MHD)  to electron scales have been recently studied in \citep{alexandr09,sahraoui09}.  In a restricted statistical study of 7 time intervals under different plasma conditions, \citet{alexandr09} show that  the spectrum appears as quasi-universal when the Kolmogorov's normalization is used (see e.g. \citep{Frisch95}). In ordinary fluids, the Kolmogorov universal function is $E(k)\ell_d/\eta^2 = (k\ell_d)^{-5/3}$, with $\eta$ being the kinematic viscosity and $\ell_d$ the dissipation scale. 
In our case, using  $\ell_d=\rho_e$ and $\eta=cst$, we got  collapsed spectra   $E(k)\rho_e \simeq (k\rho_e)^{-5/3}$ at MHD scales and $\simeq (k\rho_e)^{-2.8}$ at ion scales. 
 
The transition between these two well-defined power-laws is found in the vicinity of the ion scales where the electron fluid has a significant drift with respect to the ion fluid  and where fluctuations can be sensitive to the local plasma conditions \citep{Bale09, Bourouaine10, Matteini2011}. This ion spectral break transition is not universal \citep{smith2006,alexandr10}. 

 At electron scales, the spectral shape is not yet completely established. \citet{sahraoui09} show a clear spectral break at Doppler shifted $\rho_e$ in the electron foreshock region.  This suggests a possible cascade at scales smaller than $\rho_e$.  \citet{alexandr09} show however that in the free solar wind  a curved spectral shape is observed at these scales, instead of a break between two well defined power laws, suggesting dissipation of turbulence. 

We present here a relatively large statistical study of 102 spectra  measured by the Cluster mission~\citep{Escoubet1997} in the free solar wind.  We find that their curved spectral shapes can be fitted by a unique function $\sim k^{\alpha}\exp(-k/k_0)$ in a range starting at the vicinity of the ion break point and extending beyond electron scales. 

This result does not discard the possibility of another cascade at scales smaller than $\rho_e$, inaccessible  with present instruments, but   gives an increased strength to the hypothesis that at electron scales, i.e. at $\ell \sim 1$~km, there is  dissipation of the electromagnetic turbulence in the solar wind. Whether this dissipation is final or only partial is still an open question.

The  Cluster spacecraft was designed as a magnetospheric mission and its excursions in the solar wind not connected to the terrestrial bow-shock  are rather limited in time. This is why it is difficult to find a large number of  long time intervals to study  the  MHD inertial range. Nevertheless, time intervals of 10 minutes, as we will consider here, are frequent, and long enough to study kinetic scales.

We have selected homogeneous intervals among the first five years of the Cluster mission (2001-2005). In order to eliminate solar wind intervals when Cluster is magnetically connected to the  Earth's bow-shock, we have used (i) electrostatic wave spectrograms, which show clearly waves typical of the electron foreshock and (ii) the connection depth, calculated with straight field lines and a paraboloidal shock model \citep{Filbert1979,Lacombe1988}.  When the interplanetary magnetic field ${\bf B}$ is quasi-parallel to the solar wind velocity ${\bf V}$, Cluster is connected to the shock. Thus, our data set only contains intervals for which the angle $\Theta_{BV} $ between  ${\bf B}$ and  ${\bf V}$ is larger than $60^{\circ}$.  
If the turbulent  fluctuations have a phase speeds $V_{\phi}\ll V$,  we can detect  by Doppler shift the fluctuations with ${\bf k\| V}$. As  ${\bf B}$ and  ${\bf V}$ are quasi-perpendicular, this means that we mainly study fluctuations with ${\bf k\perp B}$. We apply the Taylor hypothesis (i.e., the direct relationship between time $\tau$ and space scales  $\ell=V\tau$) to get wave-number $k$ from frequency $f$. However, 
about $\sim 10\%$ of the pre-selected intervals show the presence of   right hand polarized whistlers in parallel propagation. For these waves  the Taylor hypothesis is not applicable, because their $V_{\phi} > V$. We  discard these intervals. This data selection process gives us 102~intervals.

In our statistical sample, the plasma conditions vary as usually in the free solar wind in fast and slow streams: bulk speed is  $V\in [300,700]$~km/s, ion temperature is $T_i\in [4,80]$~eV, temperature ratio is $T_i/T_e\in[0.3,5]$, mean magnetic field is $B\in[3,20]$~nT, ion plasma beta is $\beta_i\in[0.1,10]$, electron plasma beta is $\beta_e\in[0.1,20]$, ion and electron temperature anisotropy $T_{\perp}/T_{\|}$ are smaller than $1$.  As usual, the plasma parameters are inter-correlated:  the strongest correlation  is observed between $V$ and $T_i$ with a  correlation coefficient of  $C(V,T_i)\sim 0.8$, followed by the correlation between magnetic  and ion thermal pressure, electron and ion temperatures,  magnetic field  intensity and electron temperature, $C(B^2,nkT_i)\sim C(T_i,T_e) \sim C(B,T_e) \sim 0.6$.

The Power Spectral Density (PSD) of magnetic fluctuations as a function of frequency in the spacecraft frame $P(f)$ is obtained from the STAFF instrument measurements  \citep{Cornilleau1997}  on Cluster. This instrument has two components: (1)~The Search Coil sensors (SC), which measures magnetic waveforms at frequencies $f\in[0.1,12.5]$~Hz in normal mode, and up to $180$~Hz in burst mode; and (2)~the Spectrum Analyser (SA), which measures magnetic and electric spectra from 8~Hz to 4~kHz every 4~s  (so that for a 10 minutes interval there are 150 individual spectra). In this study we combine SC data in normal mode, and SA spectra  for the frequencies where the Signal to Noise Ratio (SNR) is larger than~3.
Note that  measurements with  SNR$\simeq 5-10$ are already affected by the instrumental noise, which becomes dominant  for SNR $\sim 3$   \citep{alexandr10}. This instrumental noise limit allows us to use SA-data up to $60-400$~Hz as a function of the turbulence intensity (i.e., for the most intense spectrum, we have valid observations up to $400$~Hz). A poor calibration of the first 3 frequencies of SA (at 8, 11 and 14~Hz) [Y. de Conchy and N. Cornilleau, private communication, 2011], was corrected by an interpolation of these points between the highest SC frequencies and the 4th and 5th points of SA spectra. The linear interpolation between $\log P(f)$ and $\log f$ is possible as far as the spectra follow a power-law at these frequencies.

\begin{figure}
\includegraphics[width=5.5cm, angle=90]{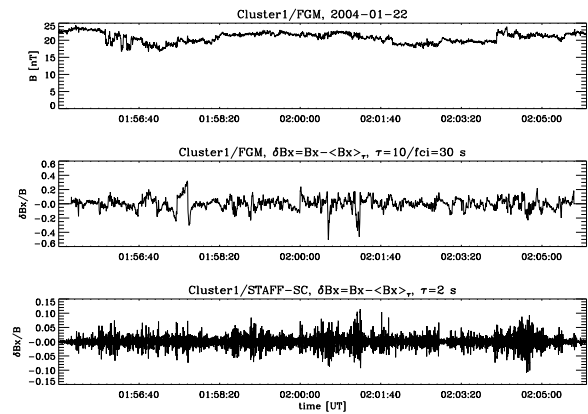}
\caption{\label{fig:Bt-case} Cluster/FGM \&  STAFF measurements in the solar wind on 22th of January 2004 (reference 10-minutes time interval, see text); (a) magnetic field modulus; (b) fluctuations $\delta B_x/B$ at scales smaller than $10/f_{cp} =30$s; (c) small scales fluctuations  $\delta B_x/B$ within the $[0.5,10]$~Hz frequency range. }
\end{figure}

Let us start with the most intense turbulent interval (called reference interval in the following), Figure~\ref{fig:Bt-case}. It lies in the free solar wind downstream of the interplanetary bow shock passed by Cluster at 01:35~UT on January 22, 2004.  The mean field is very high, around $20$~nT (top panel). In the two other panels we show  magnetic field fluctuations $\delta B_x$ 
at different scales $\tau$: (i) FGM measurements of  $\delta B_x=B_x-\langle B_x\rangle_{\tau}$  with $\tau=10/f_{ci} \simeq 30$~s are within the Kolmogorov's inertial range; (ii)  STAFF-SC data in the range  $[0.5,10]$~Hz, i.e., at frequencies higher than the ion spectral break ($\sim 0.3$~Hz). One can see that within the inertial range,  the relative amplitudes may be large, $\delta B_x/B\sim 0.5$. 

\begin{figure}
\includegraphics[width=8cm]{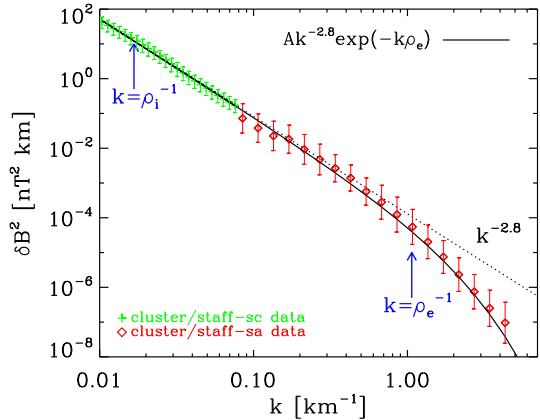}
\caption{\label{fig:spec-0} Spectrum of magnetic fluctuations at scales smaller than 1000 km, measured by Cluster-1/STAFF on 22/01/2004  (reference spectrum, see text). Green crosses represent the SC measurements, red diamond show the SA measurements. The blue arrows indicate inverse ion and electron Larmor radii.   Dotted line indicates the $k^{-2.8}$ power-law. The solid line gives the fluid-like dissipation law $Ak^{-2.8}\exp(- k\rho_e)$. }
\end{figure}

For the analyzed time period, Figure~\ref{fig:spec-0} shows the PSD of magnetic fluctuations as a function of the wave-number $P(k)=P(f)V/2\pi$, which are determined using the Taylor hypothesis ($k=2\pi f/V$) and the energy conservation law $\int P(k)dk = \int P(f)df$. Green crosses show the Morlet wavelet spectrum \citep{Torrence1998} of STAFF-SC measurements  (presented in the bottom panel of Figure~\ref{fig:Bt-case}). Red diamonds display the STAFF-SA data for the same time period. (In this plot we keep the 3 first poorly calibrated data points, one can see them around $k=0.1$~km$^{-1}$ and compare with the result of the interpolation in Figure 3). 
The error bars are estimated from the variance of the PSD at each frequency \citep{alexandr10}. This spectrum is valid up to $\simeq 400 Hz$, which gives us the maximum wave-vector $k\sim 4$~km$^{-1}$ (while $1/\rho_e \simeq 1$~km$^{-1}$).  This is the smallest scale ever measured with a good sensitivity at 1~AU in  the solar wind.

From Figure~\ref{fig:spec-0} one can see that the two instruments are in agreement and that the high-$k$ part displays a clear curvature. The curved spectrum makes one think of    the high-wave number tail found in the 3D fluid turbulent cascade (e.g., \citet{Chen93}): 
\begin{equation}\label{eq:diss}
P(k)\sim k^{\alpha}\exp(ck/k_d)
\end{equation}
where $k_d\sim 1/\ell_d$ is the dissipation wave number. 

In \citep{alexandr09} we have shown that the electron Larmor radius $\rho_e$ can play the role of a dissipation scale $\ell_d$ in the collisionless solar wind, and that the quasi-universal spectrum between ion and electron scales follows a $k^{-2.8}$ power-law. Following this line, we plot $A k^{-2.8}\exp(-k\rho_e)$ as a solid line in Figure~\ref{fig:spec-0}. Only the amplitude $A$ has been adjusted to the STAFF-SC (green) spectrum, the STAFF-SA measurements fall on this curve without any particular fitting. We plot as well the $k^{-2.8}$ power-law by a dotted line in order to underline the departure of the solar wind spectrum from the power-law shape.

\begin{figure}
\includegraphics[width=5.5cm]{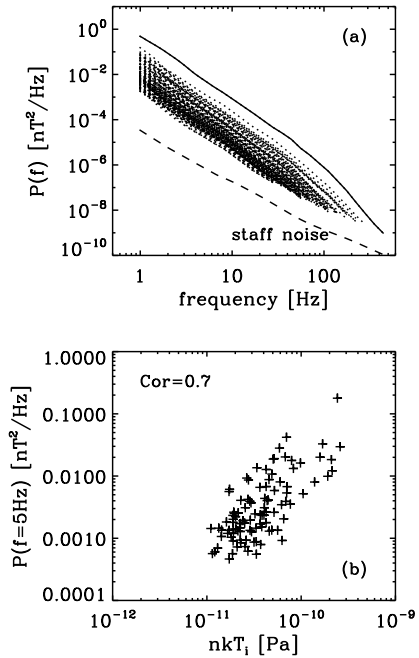}
\caption{(a) Raw 102 frequency spectra with signal to noise ratio  greater than 3 measured by
Cluster-1/STAFF in the free solar wind.  The dashed line shows the instrument noise level.
(b) PSD of magnetic fluctuations $\delta B$ at a fixed frequency as a function of the ion thermal pressure in the solar wind, correlation coefficient is 0.7.}
\label{fig:spec102}
\end{figure}

Let us now check the generality of this findings, by considering the whole set of 102 spectra, presented in Figure~\ref{fig:spec102}. The top panel shows  $P(f)$-spectra, the reference spectrum is presented by solid line, the other 101 spectra by dotted lines.  The spectra look very similar: only their amplitude changes as a function of the solar wind pressures. The best correlation of the spectral intensity is found with the ion thermal pressure $nkT_i$, Figure~\ref{fig:spec102}(b). Similar result was found at MHD scales in \citep{Grappin90}.

\begin{figure}

\includegraphics[width=8.cm]{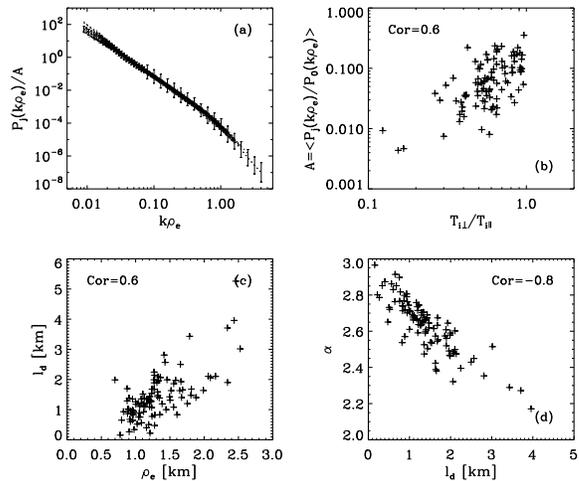}
\caption{(a)Normalized $P(k\rho_e)/A$ spectra; (b) $A$ as a function of the ion temperature anisotropy, correlation of 0.64.}
\label{fig:spec102-k}
\end{figure}

The similarity of the $P(f)$ spectra suggests  that there is a quasi-universal spectrum such that any observed spectrum can be presented as a function of this universal spectrum with an appropriate rescaling  \citep{Pedrosa99}.

Figure~\ref{fig:spec102-k}(a)  shows the superposition of rescaled spectra $\tilde{P}(k\rho_e)/A$, where $P(k)dk = \tilde{P}(k\rho_e)d(k\rho_e)$ and  $A$ is the  amplitude of the spectrum $\tilde{P}_j(k\rho_e)$ for a given period  ($j=1,...,101$) relative to that of a reference period, $\tilde{P}_0$,   $A=\langle \tilde{P}_j(k\rho_e)/\tilde{P}_0(k\rho_e)\rangle$; $\langle.\rangle$ denotes the average over the valid points of each  $\tilde{P}_j$ spectrum, but for $f$ higher than a certain limit  to avoid the range close to the ion spectral break, $f \geq 3$~Hz, corresponding to $k\rho_e \geq 0.02$. One can see that these spectra $\tilde{P}(k\rho_e)/A$ are superposed within the error-bars of the  $\tilde{P}_0(k\rho_e)$ spectrum.

The normalized spectra $\tilde{P}(k \rho_e)$   no longer display  the   dependence on the ion thermal pressure, observed for the  $P(f)$-spectra.  We find instead a new dependence, namely, on the ion temperature anisotropy, $A\sim (T_{i\perp}/T_{i\|})^{1.5}$, see Figure~\ref{fig:spec102-k}(b). The dependence between the turbulence intensity and  $T_{i\perp}/T_{i\|}$ was observed for the ion break scale \citep{Bale09,Bourouaine10}. Here we show that it keeps over smaller scales.   

In order to specify the functional dependence of the observed spectrum, we perform a three-parameter fitting  with equation~(\ref{eq:diss}), precisely with $P(k)=A_0 k^{-\alpha}\exp(-k\ell_d)$ with $A_0$, $\alpha$ and $\ell_d$ as free parameters. 
This fitting gives the scale $\ell_d$, that is correlated with $\rho_e \sim \sqrt{T_e}/B$ (see panel (c)), but not with $\lambda_e\sim 1/\sqrt{N}$ (not shown). This correlation between   $\rho_e$ and  $\ell_d$ confirms our findings  \citep{alexandr09} on the r\^ole of  $\rho_e$. The amplitude $A_0$ correlates, as expected, with $A$.

Last, the fitting process leads to a spectral index varying in the range $\alpha \in [2.2,2.9]$
showing a nice anti-correlation with the dissipation scale $\ell_d$, see Figure~\ref{fig:spec102-k}(d).
Such an anti-correlation is indeed expected from a balance between energy injection and dissipation, 
with the dissipation scale going to zero when the spectral slope approaches the value $3$
\citep{Rose-Sulem1978,Matthaeus2008}.

If we summarize the findings of the present study, the dissipation range spectrum in a collisionless space plasma can be written as
\begin{equation}\label{eq:diss-sw}
P(k)=A k^{-\alpha}\exp(- k\ell_d).
\end{equation}
 where $A$ is related to the anisotropy of ions, $\ell_d\sim \rho_e$ and $\alpha$ is a function of the dissipation length.  

This provides a unique description of the solar wind spectrum starting at ion scales and going beyond the electron scales, which is much more general than others proposed previously \citep{alexandr09,sahraoui09}.
The exponential tail of the spectrum indicates that the effective dissipation of magnetic fluctuations in the solar wind has a wave number dependence similar to that of the resistive term in collisional fluids   $\sim  \triangle \delta B \sim k^2 \delta B$.  In the literature, there are several models of kinetic range of the solar wind turbulence (see e.g. \citep{Schekochihin2009ApJS, Howes2010, Rudakov2011}). However, none of them leads to the exponential tail found here.


\bibliography{biblio11}
\end{document}